\documentclass[trackchanges, twocolumn]{aastex701}
\usepackage{comment}
\usepackage{amsmath} 
\usepackage{bm}
\usepackage{mathrsfs}
\usepackage{subcaption}
\usepackage{afterpage}
\usepackage{placeins}
\usepackage[normalem]{ulem}

\begin{document}

\title{Reconnection-driven Plasma Loading for Prompt Emission in Tidal Disruption Event Jets}

\author[orcid=0009-0000-1257-5133]{Rin Oikawa}
\affiliation{Astronomical Institute, Graduate School of Science, Tohoku University, Sendai 980-8578, Japan}
\email[show]{oikawa.rin@astr.tohoku.ac.jp} 

\author[orcid=0000-0003-2477-9146]{Yuri Sato}
\affiliation{Astronomical Institute, Graduate School of Science, Tohoku University, Sendai 980-8578, Japan}
\email[show]{yuri@astr.tohoku.ac.jp}

\begin{abstract}
A tidal disruption event (TDE) occurs when a star passes sufficiently close to a black hole (BH) to be disrupted by its tidal forces. Although several hundred TDEs have been identified to date, only a small fraction launch relativistic jets. These jets are thought to be Poynting-flux-dominated outflows powered by the Blandford-Znajek (BZ) mechanism. However, the origin of the jet plasma required to account for the observed prompt emission remains poorly understood and constitutes a key unresolved problem in the formation of relativistic TDE jets.
In this Letter, we propose magnetic reconnection near the BH as a physical mechanism for injecting plasma into the BZ jet during the super-Eddington accretion phase. Once sufficient magnetic flux accumulates near the BH, an equatorial current sheet forms, in which magnetic reconnection produces high-energy gamma rays. We demonstrate that photon-photon interactions generate a sufficient particle supply to account for the observed prompt emission. We further show that our plasma loading scenario allows the jet to remain magnetically dominated out to the prompt emission region and that magnetic dissipation can consistently reproduce the observed prompt emission properties. Furthermore, our scenario predicts MeV and GeV gamma-ray flares, providing a direct means of testing this scenario with future gamma-ray observatories.
\end{abstract}


\section{Introduction}\label{Introduction}

A star that passes sufficiently close to a supermassive black hole (SMBH) is tidally disrupted, producing a tidal disruption event (TDE) \citep{Hills1975,Rees1988,EK1989}. 
Roughly half of the stellar debris remains gravitationally bound and falls back toward the BH. 
During the early phase, the fallback rate can exceed the Eddington accretion rate, leading to the formation of a super-Eddington accretion flow. 
This accretion flow radiates brightly in the optical, ultraviolet, and X-ray bands
\citep{Loeb1997,Strubbe2009,Lodato2011,Komossa2015,MS2016,Roth2016}.
While most TDEs are dominated by thermal emission from the accretion flow, a small fraction produces powerful relativistic jets.

Some observed jetted TDEs include Swift J1644+57~\citep{Bloom2011,Burrows2011}, Swift J2058+05~\citep{Cenko2012}, Swift J1112-8238~\citep{Brown2015}, and AT~2022cmc\citep{Andreoni2022}.
These events exhibit prompt nonthermal X-ray emission with rapid variability on timescales of $\delta t_{\rm obs}\sim10^2$-$10^3\,{\rm s}$ and isotropic-equivalent luminosities of $L_{X,\rm iso}\sim10^{47}$-$10^{48}\,{\rm erg\,s^{-1}}$ \citep{Bloom2011,Burrows2011,Cenko2012,Brown2015,Andreoni2022}.
The physical origin of these relativistic jets remains an open question.

One possible mechanism for powering relativistic jets in TDEs is the Blandford–Znajek (BZ) mechanism, which extracts rotational energy from the BH through magnetic field lines threading its event horizon \citep[e.g.,][]{Blandford1977,DeColle2020,Sato2026,Tripto2026}. 
The origin and supply of the magnetic flux required to power a BZ jet have been extensively discussed in previous studies \citep[e.g.,][]{Krolik2012,Tchekhovskoy2014,Kelley2014,Piran2015,Dai2018,Curd2019,Abolmasov2026,Tripto2026}. However, the origin of the plasma carried by the jet, particularly the particles responsible for the prompt emission, remains poorly understood because charged particles from the surrounding accretion flow or wind cannot readily diffuse across magnetic field lines into the electromagnetically dominated jet region.
The plasma content and composition play crucial roles in determining both the radiative and dynamical properties of BH jets.\footnote{The jet plasma also carries the electric currents required to generate the toroidal magnetic field and the associated outward Poynting flux \citep{Blandford1977}.} Understanding the physical mechanism that supplies plasma to the jet is a key unresolved problem in BH jet formation.

\begin{figure}
\includegraphics[width=0.48\textwidth]{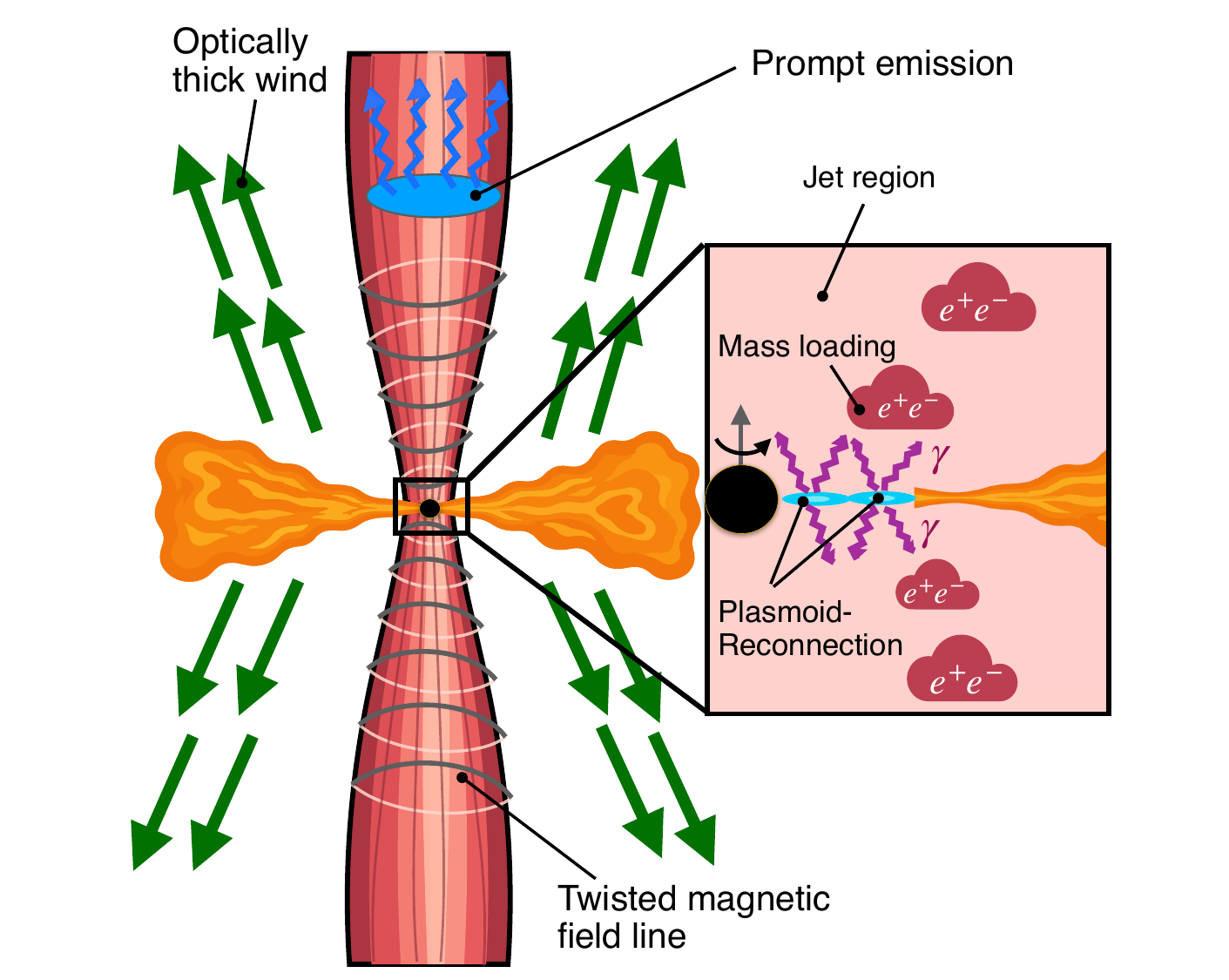}
\vspace{-0.05cm}
\caption{
Schematic picture of plasma loading by magnetic reconnection for jetted TDEs. As magnetic flux accumulates near the BH, a magnetically dominated BH magnetosphere forms. The magnetic pressure compresses the inner accretion flow toward the equatorial plane, forming an equatorial current sheet. Magnetic reconnection in the current sheet produces high-energy gamma rays, which generate electron--positron pairs through photon--photon pair production in the magnetosphere. The pair-loaded magnetosphere launches a BZ jet, which is subsequently accelerated by the Lorentz force and produces prompt emission through magnetic dissipation at larger radii.
}
\label{fig:theory picture}
\end{figure}

In this Letter, we propose magnetic reconnection during the super-Eddington phase of a TDE as a physical mechanism for plasma loading into a BZ jet. 
Recent high-resolution general relativistic magnetohydrodynamic (GRMHD) simulations have shown that magnetic reconnection can occur in the vicinity of BHs \citep{Ripperda2022}. Motivated by these results, we investigate whether high-energy gamma rays produced by magnetic reconnection near the BH can generate sufficient electron–positron pairs through photon–photon interactions to load the BZ jet with plasma (see Figure~\ref{fig:theory picture}). We analytically demonstrate that the resulting pair supply is sufficient to account for the observed prompt emission. Our plasma loading scenario further allows the jet to remain magnetically dominated out to the prompt emission region, where magnetic dissipation can reproduce the observed prompt emission properties. Finally, our scenario predicts two distinct and testable high-energy signatures: an MeV gamma-ray flare produced by magnetic reconnection near the BH prior to the prompt emission, and a GeV gamma-ray flare produced by inverse-Compton scattering in the prompt emission region. These signatures provide direct observational tests of our reconnection-driven plasma loading scenario with future gamma-ray observatories.

\section{Constraining the Physical Parameters of a TDE Jet from Observations}

Jetted TDEs exhibit bright and rapidly variable prompt X-ray emission in the $0.3-150\,{\rm keV}$ band, with variability timescales of $\delta t_{\rm obs}\sim10^2-10^3\,{\rm s}$ \citep{Bloom2011,Burrows2011,Cenko2012,Brown2015,Andreoni2022}. The observed isotropic X-ray luminosity reaches
$L_{X,\rm iso}\sim10^{47}$--$10^{48}\,{\rm erg\,s^{-1}}$,
corresponding to an intrinsic jet luminosity of
$L_{X,\rm jet}\approx L_{X,{\rm iso}}\times(\theta_j^2/2)\sim10^{44}$--$10^{46}\,(\theta_j/0.1\,{\rm rad})^2\,{\rm erg\,s^{-1}}$, where $\theta_j$ is the half opening angle of the jet.
The characteristic emission radius inferred from the variability timescale,
$r_{\rm em}\sim2\Gamma_j^2c\delta t_{\rm obs}$, is
\begin{equation}
\frac{r_{\rm em}}{r_g}
\sim
4.1\times10^{3}
\left(\frac{M_{\rm BH}}{10^6\,M_{\odot}}\right)^{-1}
\left(\frac{\Gamma_j}{10}\right)^2
\left(\frac{\delta t_{\rm obs}}{100\,{\rm s}}\right), \label{Observed radius}
\end{equation}
where $r_g=GM_{\rm BH}/c^2$ is the gravitational radius, $M_{\rm BH}$ is the BH mass, $G$ is the gravitational constant, and $c$ is the speed of light, $\Gamma_j$ is the bulk Lorentz factor of the jet. In this section, we first estimate the magnetic-field strength required for a BZ jet and then determine the minimum number of radiating particles required by the prompt emission.

\subsection{Magnetic fields at the jet base}

We first estimate the magnetic field strength required to power the TDE jet through the BZ mechanism. The BZ jet power is given by \citet{Tchekhovskoy2011,Kelley2014}
\begin{equation}
L_{\rm BZ}=\frac{c\xi_{\rm BZ}}{2\pi^2r_H^2}\Phi^2,
\end{equation}
where $\xi_{\rm BZ}$ is the effective BZ efficiency, $\Phi=2\pi r_H^2B_H$ is the magnetic flux threading the BH horizon, $B_H$ is the magnetic field strength at the horizon, $r_H=(1+\sqrt{1-a_{\rm spin}^2})r_g=x_Hr_g$ is the horizon radius, $a_{\rm spin}$ is the dimensionless BH spin parameter.
We consider a rapidly spinning BH with $a_{\rm spin}\sim1$, for which $r_H\sim r_g$. 
We assume that the observed jet X-ray emission satisfies $L_{X,\rm jet}\lesssim L_{\rm BZ}$, 
giving the magnetic field strength as
\begin{eqnarray} 
B_H&\gtrsim&\left(\frac{L_{X,\rm jet}c^3}{2G^2\xi_{\rm BZ}x_H^2M_{\rm BH}^2}\right)^{1/2}\notag\\ &\sim&2.7\times10^6 \left(\frac{L_{X,\rm jet}}{10^{45}\,{\rm erg\,s^{-1}}}\right)^{1/2}\left(\frac{M_{\rm BH}}{10^6\,M_\odot}\right)^{-1}\notag\\ &&
\times \left(\frac{\xi_{\rm BZ}}{0.1}\right)^{-1/2}x_H^{-1}\,{\rm G}. \end{eqnarray}
Therefore, a horizon magnetic-field strength of at least order $10^6$~G is required for the BZ mechanism to account for the observed jet X-ray luminosity.

How a strong magnetic field can be accumulated on the BH following the stellar disruption is a major challenge.
Several theoretical works have proposed mechanisms for supplying and transporting the magnetic flux required to launch relativistic jets in TDEs. One possibility is that the flux is supplied directly by the disrupted star through its preexisting magnetic field \citep{Kelley2014}.
The magnetic flux inferred from observed stellar magnetic-field strengths is generally insufficient to power the observed jets, indicating that an additional amplification mechanism is required. 
\citet{Abolmasov2026} demonstrated that a small-scale turbulent dynamo can amplify the stellar seed field during debris circularization, although whether these fields can form a coherent, horizon-threading flux bundle remains unclear.
Another possibility is the Lasso mechanism, in which magnetic flux at large radii is captured and transported toward the BH by the TDE debris stream \citep{Tchekhovskoy2014,Kelley2014}. 
A preexisting accretion disk may also supply magnetic flux through interaction with the debris stream \citep{Chan2019,Curd2025}. 
Alternatively, the magnetic flux may be generated within a compact accretion flow formed after the disruption \citep{Krolik2012,Piran2015,Dai2018,Curd2019}. 
More recently, \citet{Tripto2026} considered a preexisting magnetized accretion disk with a magnetic field strength of $\sim10^4\,{\rm G}$. Despite these proposed scenarios, it remains unclear which mechanism is the most plausible. In this Letter, we do not specify the origin of the magnetic flux and assume that sufficient magnetic flux to power the jet has accumulated on the horizon through some physical process.

\subsection{Plasma content required for prompt emission}
\label{sec:estimate plasma number}

Even if sufficient magnetic flux is available to power the jet, the BZ outflow must also contain enough particles to produce the observed prompt emission.
We estimate the minimum number of radiating particles required by the prompt X-ray luminosity.
For this estimate, we assume that the prompt X-ray emission is produced by synchrotron radiation\footnote{Photospheric emission has also been considered as one of the possible mechanisms of the prompt emission \citep[e.g.,][]{Crumley2016}. Since estimates of the number of particles required in the jet depend on the assumed radiation mechanism, Eq.~\eqref{Required number particles} cannot be directly applied if the prompt emission is photospheric. We discuss the case of photospheric emission in Section~\ref{Sec:summary}.}. The synchrotron power radiated by a single electron can be expressed in terms of the characteristic synchrotron photon energy $E_{p}$ and the magnetic-field strength $B$, as
\begin{equation}
P_{\rm syn}=\frac{\sigma_T m_e c^2}{3he}
E_{p}B,
\end{equation}
where $\sigma_T$ is the Thomson cross section, $m_e$ is the electron mass, $h$ is the Planck constant, and $e$ is the elementary charge.
In the BZ mechanism, the rotation of the BH winds up the magnetic field lines threading the horizon, generating a toroidal magnetic field that supports the outward Poynting flux. Far from the BH, the toroidal component dominates and decreases approximately as $B_j\approx B_H(r/r_g)^{-1}$ \citep{Komissarov2009,Toma2013}. The magnetic-field strength at the prompt emission radius is therefore estimated as $B_j\sim 10^{3}\, {\rm G}$.

The minimum number of radiating particles required in the prompt emission region is then
\begin{eqnarray}
N_{\rm req}
&\simeq&
\frac{L_{X,\rm jet}}{P_{\rm syn}}
\notag\ \\
&\sim&
1.1\times10^{45}
\left(
\frac{L_{X,\rm jet}}
{10^{45}\,{\rm erg\,s^{-1}}}
\right)
\left(
\frac{B_j}{10^3\,{\rm G}}
\right)^{-1}
\left(
\frac{E_{p}}{10\,{\rm keV}}
\right)^{-1}.
\label{Required number particles}
\end{eqnarray}
Assuming that the radiating particles are advected from the jet base to the prompt-emission region without substantial losses, at least $\sim10^{45}$ particles are required at the jet base. The origin of this plasma supply is a key unresolved problem. In the following section, we investigate whether magnetic reconnection near the BH can generate sufficient electron–positron pairs to load the jet and satisfy this requirement.

\section{Plasma loading at the jet base}
\label{sec:Plasma loading}

Charged particles in the surrounding accretion flow or wind cannot readily diffuse across magnetic field lines into the jet funnel because BZ-powered jets are electromagnetically dominated. One possible channel for supplying plasma to the jet is through interactions involving neutral particles. In active galactic nucleus (AGN) jets, plasma loading through photon–photon interactions involving MeV gamma rays produced by radiatively inefficient accretion flows has been widely discussed \citep{Moscibrodzka2011,Wong2021}. However, during the early phase of a TDE, the accretion flow is optically thick super-Eddington and its thermal emission alone may not efficiently produce the MeV gamma rays required for plasma loading\footnote{\citet{Chan2026} demonstrate that multiple Compton scattering of thermal disk photons produces a Wien tail that generates electron--positron pairs, which are subsequently advected by the bulk flow into the jet funnel in the standard and normal evolution (SANE) regime. Whether this mechanism operates efficiently in the magnetically arrested disk (MAD) regime remains uncertain.}.

Recent high resolution GRMHD simulations suggest that the magnetic pressure of the BH magnetosphere compresses the inner accretion flow toward the equatorial plane, leading to the formation of an equatorial current sheet in which magnetic reconnection can occur when a strong magnetic flux accumulates near the BH \citep{Ripperda2022}. We propose a plasma loading scenario in which magnetic reconnection accelerates particles, produces MeV gamma rays, and thereby supplies plasma to the jet. In this section, we first show that an equatorial current sheet can form for the horizon magnetic-field strength required to account for the observed jet luminosity, $B_H\sim10^{6}\,{\rm G}$, and then estimate the amount of plasma supplied by pair production associated with the MeV gamma-ray flare from magnetic reconnection.

\subsection{Triggering the Magnetic Reconnection Near the Black Hole}

We examine whether the accumulated magnetic flux can compress the super-Eddington accretion flow toward the equatorial plane strongly enough to form a current sheet. 
A super-Eddington accretion flow is expected to be supported primarily by radiation pressure.
The radiation pressure in the flow can be written as
\begin{equation}
P_{\rm rad}=\frac{1}{3}aT^{4}
\sim2.5\times10^{9}\,
\left(\frac{T}{10^6\,{\rm K}}\right)^4\,{\rm erg\:cm^{-3}},
\end{equation}
where $a = 7.57\times10^{-15}\,{\rm \,cm^{-3}\,K^{-4}}$ is the radiation constant, and $T$ is the disk temperature. When a BH magnetosphere forms, its magnetic pressure also contributes to the vertical force balance of the accretion flow.
Magnetic pressure in the BH magnetosphere can be written as
\begin{equation}
P_{\rm mag}=\frac{B_{H}^2}{8\pi}
\sim4.0\times10^{10}\,
\left(\frac{B_{H}}{10^6\,{\rm G}}\right)^2\,{\rm erg\:cm^{-3}}.
\end{equation}
 The disk scale height is expected to be determined by hydrostatic balance between the radiation pressure and the magnetic pressure:
\begin{equation}
    \frac{P_{\rm rad}}{H} \approx \frac{2P_{\rm mag}}{R},
\end{equation}
where $R$ is the radial direction. The factor of 2 in the second term accounts for the magnetic pressure exerted on both the upper and lower surfaces of the disk. The disk scale height near the horizon is estimated to be
\begin{equation}
    \frac{H}{R}\sim 0.01\,\left(\frac{T}{10^6\,{\rm K}}\right)^{4}\left(\frac{B_H}{10^{6}\,{\rm G}}\right)^{-2}. \label{scale height}
\end{equation}
As the disk becomes geometrically thin near the horizon, the oppositely directed magnetic field lines above and below the disk are brought into close contact, leading to the formation of an equatorial current sheet. The Lundquist number of the resulting current sheet is defined as $S={l_{\rm rec}v_A}{/\eta_{\rm m}}$,
where $l_{\rm rec}=f_l r_g$ is the current sheet length and $f_l$ is a dimensionless parameter, $v_A$ is the Alfvén velocity, and $\eta_{\rm m}$ is the magnetic diffusivity, and is expected to be very large. In resistive MHD, a Sweet--Parker current sheet becomes unstable to the plasmoid instability when $S$ exceeds the critical value $S_{\rm c}\sim10^4$ \citep{Loureiro2007,Bhattacharjee2009,Huang2013}. This condition corresponds to a current-sheet half-thickness $\delta_{\rm SP}$ satisfying
\begin{equation}
\frac{\delta_{\rm SP}}{l_{\rm rec}}\sim S^{-1/2}
\lesssim S_{\rm c}^{-1/2}\sim0.01. \label{Plasmpid condition}
\end{equation}
From Eqs.~\eqref{scale height} and \eqref{Plasmpid condition}, we find that the equatorial current sheet satisfies the condition for plasmoid instability and is therefore expected to undergo plasmoid-mediated magnetic reconnection \citep{Beloborodov2017,Ripperda2022}.

\subsection{High Energy Photons from Magnetic Reconnection}
\label{sec:radiation_reconnection}

\begin{figure}
\centering
\includegraphics[width=0.45\textwidth]{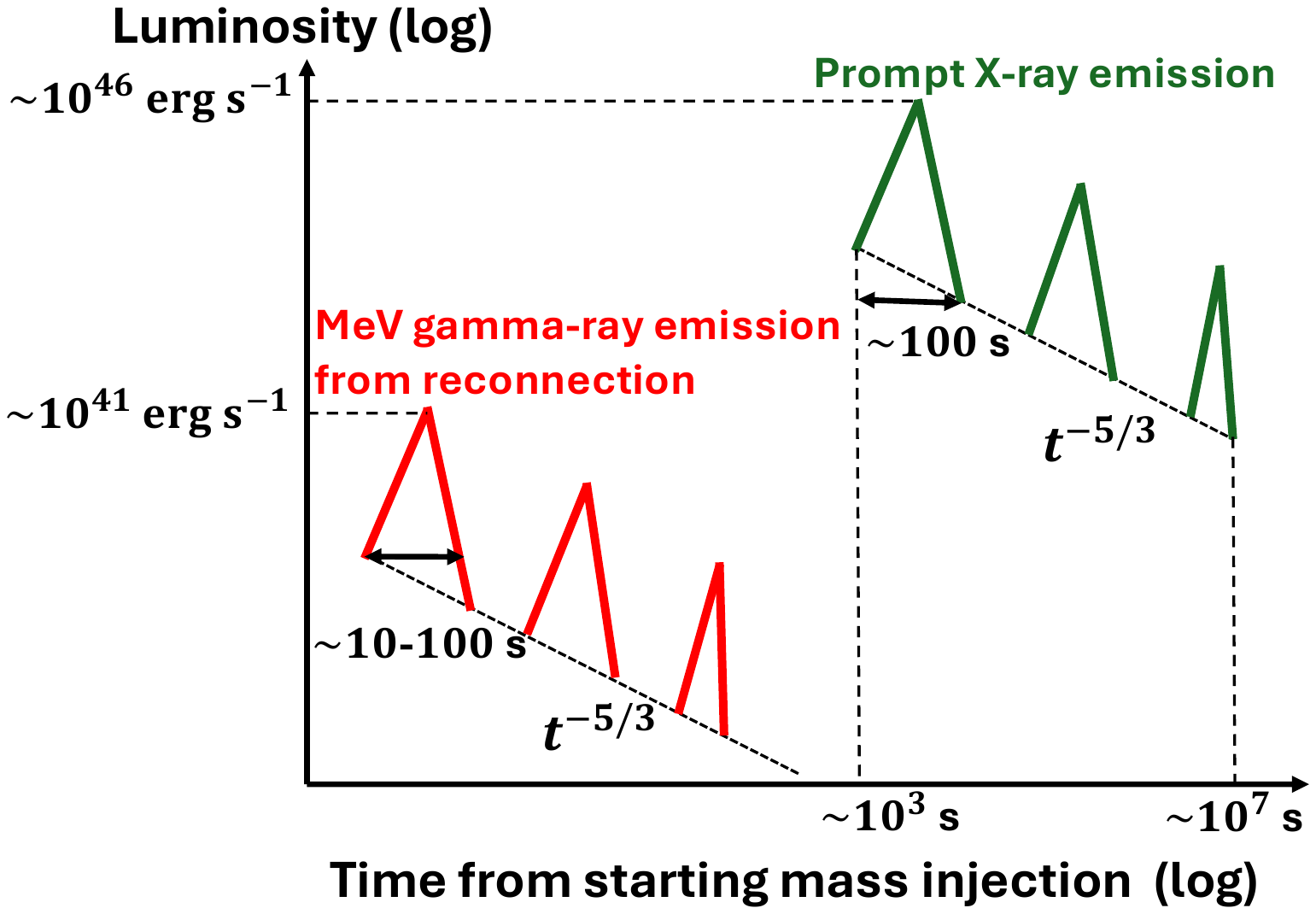}
\vspace{-0.05cm}
\caption{
Schematic view of the light curves in our scenario for a BH with a mass of $M_{\rm BH}=10^{6}M_{\odot}$.
The red and green curves represent the MeV gamma-ray emission from magnetic reconnection near the BH and the X-ray prompt emission from the jet, respectively. Both components decay as $t^{-5/3}$ because their luminosities scale as $B^2\propto\dot{M}\propto t^{-5/3}$.
The prompt emission is expected to occur approximately $r_{\rm em}/c$ after plasma injection begins (see Eq.~\ref{Observed radius}).
}
\label{fig:lightcurve}
\end{figure}

A current sheet separating two oppositely directed magnetic fields becomes unstable to the tearing mode, leading to the formation of a chain of plasmoids separated by magnetic X-points \citep{Uzdensky2010}. Particles are accelerated by the reconnection electric field $E_{\rm rec}\sim \beta_{\rm rec}B$ near the X-points, where $\beta_{\rm rec}$ is the reconnection rate. In the plasmoid-mediated fast reconnection regime it is typically given by \citep{Bhattacharjee2009,Uzdensky2010,Guo2024,Sironi2025}
\begin{equation}
\beta_{\rm rec} \sim
\begin{cases}
0.01 & \text{(collisional regime)},\\
0.1  & \text{(collisionless regime)}.
\end{cases}
\end{equation}
In the absence of cooling effects, the particle energy gain can reach $\sim \sigma m c^2$, where $\sigma$ is the magnetization parameter defined as
\begin{equation}
\sigma\equiv\frac{B^2}{4\pi n m c^2}, \label{definition magnetization parameter}
\end{equation}
where $m$ is the rest mass of a plasma particle, and $n$ is the plasma number density.
The acceleration timescale of electrons at the X-points can be written as $t_{\rm acc}\approx{\gamma_e m_e c}/({eB_H\beta_{\rm rec}}$), where $\gamma_e$ is the electron Lorentz factor. The accelerated electrons cool via synchrotron radiation on a timescale of $t_{\rm syn}\approx{6\pi m_e c}/({\sigma_T B_H^2\gamma_e})$. Equating the acceleration time and cooling time, we can obtain maximum Lorentz factor as
\begin{equation}
    \gamma_{e,\rm{max}}\approx\sqrt{\frac{6\pi e\beta_{\rm rec}}{\sigma_T B_H}}\sim 1.2\times10^{4}\, \left(\frac{\beta_{\rm rec}}{0.01}\right)^{\frac{1}{2}}\left(\frac{B_H}{10^6\,{\rm G}}\right)^{-\frac{1}{2}}. \label{Maximum Lorentz factor}
\end{equation}
The synchrotron photon energy for these electrons can be written as
\begin{equation}
    E_{\gamma,\rm max}\approx\frac{heB_H\gamma_{e,\rm max}^2}{2\pi m_e c}=\frac{9 m_ec^2} {4\alpha_f}\beta_{\rm rec}\sim 1.6\,\left(\frac{\beta_{\rm rec}}{0.01}\right)\,{\rm MeV},
    \label{eq:E_syn}
\end{equation}
where $\alpha_f$ is the fine structure constant.
Magnetic reconnection in the vicinity of a BH releases a large amount of magnetic energy,
\begin{equation}
\begin{aligned}
L_{\rm rec}
    &\approx 2l_{\rm rec}^2 \beta_{\rm rec} c \frac{B_H^2}{8\pi} \\
    &\sim 5\times10^{41}\,f_l^2 \left(\frac{\beta_{\rm rec}}{0.01}\right)\left(\frac{M_{\rm BH}}{10^{6}\,M_{\odot}}\right)^2 \left(\frac{B_H}{10^6\,\rm G}\right)^2 {\rm erg\, s^{-1}}.
\end{aligned}
\label{eq:L_rec}
\end{equation}
The duration of the reconnection event
can be estimated to be \citep{Sironi2025}
\begin{equation}
    t_{\rm dur} = \frac{l_{\rm rec}}{\beta_{\rm rec}c}\sim 4.9\times 10^2\,f_l \left(\frac{M_{\rm BH}}{10^6 \,M_{\odot}}\right)
    \left(\frac{\beta_{\rm rec}}{0.01}\right)^{-1}{\rm s}.
\end{equation}
The number density of MeV gamma-ray photons produced by magnetic reconnection is roughly estimated as
\begin{equation}
n_{\gamma}\approx \frac{L_{\rm rec}}{4\pi l_{\rm rec}^2 c E_{\gamma,\rm max}}\sim 2.5\times10^{13}\,\left(\frac{B_H}{10^6\,{\rm G}}\right)^2\,{\rm cm^{-3}}.
\end{equation}
Therefore, magnetic reconnection in the vicinity of the BH can efficiently produce MeV gamma-ray photons. 
The magnetic-field strength at the jet base scales with the mass accretion rate as $B_H\propto\dot{M}^{1/2}$ \citep{Tchekhovskoy2011,Tchekhovskoy2014}, while the fallback rate evolves as $\dot{M}\propto t^{-5/3}$ \citep{Rees1988,EK1989}. Consequently, the reconnection luminosity evolves as $L_{\rm rec}\propto B_H^2\propto\dot{M}\propto t^{-5/3}$, resulting in the reconnection flare light curve illustrated in Figure~\ref{fig:lightcurve}. These photons can subsequently generate electron--positron pairs through the Breit--Wheeler process ($\gamma\gamma\rightarrow e^{+}e^{-}$), thereby supplying plasma to the BH magnetosphere.

\subsection{Mass Loading into the BZ Jets}
The MeV photons produced by reconnection can provide the seed photons for electron–positron pair production. 
In low-luminosity AGN jets, such reconnection-driven pair production has been proposed as a viable source of jet plasma \citep{Kimura2022,Oikawa2026}. 
We investigate whether the same mechanism can efficiently load the BH magnetosphere in TDEs.

The optical depth for the Breit--Wheeler process is estimated to be
\begin{equation}  \tau_{\gamma\gamma}=n_{\gamma}\sigma_{\gamma\gamma}l_{\rm rec}\sim2.5\times10^{-1}\,f_{l}\, \left(\frac{B_H}{10^6\,{\rm G}}\right)^2\left(\frac{M_{\rm BH}}{10^6\,M_{\odot}}\right), \label{optical depth}
\end{equation}
where $\sigma_{\gamma\gamma}\approx f_{\gamma\gamma}\sigma_T$ is the approximate cross section for the Breit--Wheeler process, and $f_{\gamma\gamma}\sim0.1$ \citep{Svensson1987}. Most of the MeV gamma rays produced by magnetic reconnection escape from the reconnection region, but a non-negligible fraction undergoes electron--positron pair production in the BH magnetosphere. The electron--positron pair production rate due to $\gamma\gamma$ interaction is given by
\begin{equation}
\dot{n}_{\pm}
\approx
n_{\gamma}^{2}\sigma_{\gamma\gamma}c
\sim
1.3\times10^{12}
\left(\frac{B_H}{10^6\,{\rm G}}\right)^4
\,{\rm cm^{-3}\,s^{-1}}.
\end{equation}
We obtain the injected pair number density as $n_{\pm}\approx \dot{n}_{\pm}\times(l_{\rm rec}/c)$, yielding a high pair multiplicity,
\begin{equation}
\kappa_{\pm}=\frac{n_{\pm}}{n_{\rm GJ}}
\sim 1.1\times10^{10}\,f_l\,a_{\rm spin}\,x_H^{-1}
\left(\frac{M_{\rm BH}}{10^{6}\,M_{\odot}}\right)^2
\left(\frac{B_H}{10^6\,{\rm G}}\right)^3,
\end{equation}
where
\begin{equation}
\begin{split}
n_{\rm GJ}
&=\frac{\Omega_F B_H}{2\pi e c}
\approx \frac{a_{\rm spin}B_H}{8\pi e r_H} \\
&\sim 5.6\times10^2\,a_{\rm spin}x_H^{-1}
\left(\frac{B_H}{10^6\,{\rm G}}\right)
\left(\frac{M_{\rm BH}}{10^6\,M_{\odot}}\right)^{-1}
\,{\rm cm^{-3}}
\end{split}
\end{equation}
is the Goldreich--Julian number density which characterizes the plasma density required to support the electric currents associated with the BZ process \citep{Goldreigh1969}. Here, $\Omega_F=a_{\rm spin}c/(4r_H)$ is the angular velocity of the magnetic field lines threading the horizon of a Kerr BH \citep{Blandford1977,Tchekhovskoy2010}. Since $\kappa_{\pm}\gg1$, the resulting pair density is sufficient to sustain the electric currents for the BZ process. The number of injected particles during a magnetic reconnection event is approximately given by 
\begin{equation}
N_{\rm inj}\approx n_{\pm}r_g^3\sim2.0\times10^{46}\,f_l\left(\frac{M_{\rm BH}}{10^{6}\,M_{\odot}}\right)^4\left(\frac{B_H}{10^6\,{\rm G}}\right)^4.
\end{equation}
This value exceeds the required particle number inferred from the observed prompt emission (see Eq.~\eqref{Required number particles}), demonstrating that our plasma loading scenario can explain the origin of the jet plasma.

The characteristic Coulomb mean free path of the injected pairs is estimated to be
\begin{equation}
l_{\rm c}\approx\frac{1}{n_{\pm}\pi r_e^2}
\sim6.5\times10^{11}\,f_l^{-1}\left(\frac{B_H}{10^6\,\rm G}\right)^{-4}\left(\frac{M_{\rm BH}}{10^6\,M_{\odot}}\right)^{-1}\,{\rm cm},
\label{collision length scale}
\end{equation}
where $r_e$ is the classical electron radius. 
The Coulomb mean free path is comparable to the size of the reconnection region, suggesting that both collisional and collisionless regimes may be relevant. We therefore consider $0.01 \lesssim \beta_{\rm rec} \lesssim 0.1$.

\section{Implications for Prompt emission}

We investigate the implications of our plasma loading scenario for the prompt emission mechanism of jetted TDEs. 
We first estimate the physical conditions in the prompt emission region predicted by our scenario and then discuss the radiation mechanism based on these estimates.

\subsection{Physical Conditions at the Emission Region}
The mass loading rate at the jet base can be estimated as
\begin{equation}
\dot{M}_j \approx 2m_e\dot{N}_{\pm}
\sim 7.4\times10^{18}
\left(\frac{B_H}{10^6\,{\rm G}}\right)^4
\left(\frac{M_{\rm BH}}{10^6\,M_{\odot}}\right)^3
\,{\rm g\,s^{-1}},
\end{equation}
where the total pair injection rate is roughly given by $\dot{N}_{\pm}=\int \dot{n}_{\pm}\,dV
\approx \dot{n}_{\pm}r_g^3$. 
The jet energy flux can be written as
\begin{equation}
L_j \approx L_{\rm EM}+\Gamma_j \dot{M}_j c^2, \label{energy conservation}
\end{equation}
where $L_{\rm EM}\approx cB_j^2r^2$ is the Poynting luminosity and $B_j$ is the magnetic field strength in the jet. 
At the jet base, whose quantities are denoted by the subscript $0$, the bulk Lorentz factor is expected to be $\Gamma_{j,0}\sim1$, and the electromagnetic power supplied by the BZ process dominates the kinetic energy flux $L_{\rm EM,0}\approx L_{\rm BZ}\gg\dot{M}_{j}c^2$.
Assuming a steady jet flow, we have $L_j \approx L_{\rm BZ}$ and Eq.~\eqref{energy conservation} can be rewritten as
\begin{equation}
L_{\rm BZ} \approx \Gamma_j (1+\sigma_j)\dot{M}_j c^2\approx cr^2B_j^2\left(\frac{1+\sigma_j}{\sigma_j}\right),
\end{equation}
where $\sigma_j=L_{\rm EM}/(\Gamma_j\dot{M}c^2)$ is the jet magnetization parameter, defined consistently with Equation~\eqref{definition magnetization parameter}.
The comoving magnetic field at the emission region is estimated to be
\begin{equation}
\begin{aligned}
B_j'
&\approx\left(\frac{L_{\rm BZ}}{\Gamma_j^2 r^2 c}\frac{\sigma_j}{1+\sigma_j}\right)^{1/2} \\
&\sim1.2\times10^2\,x_H\,\left(\frac{\sigma_j}{1+\sigma_j}\right)^{1/2} 
 \left(\frac{\xi_{\rm BZ}}{0.1}\right)^{1/2}\left(\frac{M_{\rm BH}}{10^6\,M_{\odot}}\right) \\
  &\:\:\:\:\:\times\left(\frac{B_H}{10^6\,{\rm G}}\right)
   \left(\frac{\Gamma_j}{10}\right)^{-1}\left(\frac{r}{10^{14}\,{\rm cm}}\right)^{-1}\,{\rm G},
\end{aligned}
\end{equation}
where we have used $B_j\approx\Gamma_jB_j'$.
The jet magnetization parameter at the prompt emission region can be approximated as
\begin{equation}
\begin{aligned}
1+\sigma_j &= \frac{L_{\rm BZ}}{\Gamma_j\dot{M}_jc^2}\sim 1.5\times10^{4}\,x_H^2 \left(\frac{\xi_{\rm BZ}}{0.1}\right)\left(\frac{\Gamma_j}{10}\right)^{-1}\\
&\:\:\:\:\:\:\:\:\:\:\:\:\:\:\:\:\:\:\:\:\:\:\:\:\:\:\:\:\:\times\left(\frac{B_H}{10^6\,\rm G}\right)^{-2}\left(\frac{M_{\rm BH}}{10^6\,M_{\odot}}\right)^{-1}.
\end{aligned}
\end{equation}
Our plasma-loading scenario suggests that the jet remains highly magnetized out to the prompt emission region. In general, internal shocks, recollimation shocks, and magnetic reconnection have been proposed as possible dissipation mechanisms responsible for prompt emission from relativistic jets \citep{Rees1994,Sironi2015,BarniolDuran2017}. In magnetically dominated relativistic flows, particularly for shocks with superluminal magnetic-field configurations, nonthermal particle acceleration is strongly suppressed, making it difficult for shocks to efficiently produce the observed nonthermal emission \citep{Sironi2009,Sironi2013,Sironi2015}. By contrast, in highly magnetized plasma, magnetic reconnection can efficiently convert magnetic energy directly into nonthermal particle energy \citep{Sironi2014,Guo2014,Sironi2015}. Therefore, we consider magnetic reconnection within the highly magnetized jet as a possible mechanism for producing the prompt emission.

\subsection{Triggering the Dissipation of Magnetic Energy }

In our model, the jet remains magnetically dominated in the prompt emission region. Magnetic reconnection provides a mechanism for converting the ordered magnetic energy into nonthermal particle energy. A possible trigger for reconnection current-sheet formation in the jet is the current-driven instability (CDI) kink mode \citep{Lyubarskii1999,Appl2000,Giannios2006,Mizuno2012,Bromberg2016,Kim2017}. Here, we estimate the characteristic radius beyond which the CDI can grow and show that it is consistent with the emission radius inferred from observations.

In the jet comoving frame, the characteristic growth timescale of the CDI can be written as \citep{Sobacchi2017}
\begin{equation}
t'_{\rm CDI}\sim \kappa \frac{R_j(r)}{c},
\end{equation}
where $\kappa$ is a dimensionless factor that depends on the magnetic-field configuration, and 
\begin{equation}
R_j(r)\approx r_H\left(\frac{r}{r_H}\right)^\alpha,
\end{equation}
is the cylindrical radius of the jet at a distance $r$ from the BH. 
The competing timescale is the adiabatic expansion timescale of the toroidal magnetic field in the jet.
The toroidal magnetic-field component is expected to dominate sufficiently far from the BH in an MHD jet \citep{Komissarov2009,Toma2013}.
Since magnetic-flux conservation gives
$B_{j,\phi} \propto R_j^{-1}\propto r^{-\alpha}$,
the adiabatic expansion timescale in the jet comoving frame is estimated as \citep{Moll2008}
\begin{equation}
t'_{\rm ad}
\sim
\frac{1}{\Gamma_j}
\frac{B_{j,\phi}}{\left|dB_{j,\phi}/dt\right|}
\sim
\frac{r}{\alpha \Gamma_j v_{j,r}},
\end{equation}
where $v_{j,r}=dr/dt$.
The CDI can grow when $t'_{\rm CDI}\lesssim t'_{\rm ad}$.
Assuming $v_{j,r}\simeq c$, 
this condition yields
\begin{equation}
\frac{r}{r_H}
\gtrsim
\left(\kappa\alpha\Gamma_j\right)^{1/(1-\alpha)}.
\end{equation}
For a parabolic jet with $\alpha=1/2$, together with $\kappa\sim10$ \citep{Nalewajko2026} and $\Gamma_j\sim10$, we obtain
\begin{equation}
r_{\rm CDI}
\sim
{\rm few}\times10^3 r_H.
\end{equation}
Thus, the CDI can become dynamically important at $r\gtrsim r_{\rm CDI}$, where its nonlinear development may distort the ordered toroidal magnetic field and generate current sheets. The resulting dissipation radius is broadly consistent with the prompt emission radius inferred from observations (see Equation~\eqref{Observed radius}). 
At the dissipation radius, the characteristic timescales satisfy $t'_{\rm dyn}<t'_{\rm CDI}\lesssim t'_{\rm ad}$. The CDI can therefore grow and distort the ordered magnetic field without disrupting the global jet propagation.

\subsection{Prompt Emission}
In our scenario, the prompt emission can be powered by the dissipation of magnetic energy through processes such as magnetic reconnection. In the dissipation region, particles are accelerated by magnetic reconnection to a characteristic electron Lorentz factor of $\gamma_e \sim \sigma_j$. These accelerated particles subsequently radiate via synchrotron cooling, producing the prompt emission. The characteristic energy of the synchrotron photons is given by
\begin{equation}
\begin{aligned}
E_{p}&\approx \frac{\Gamma_j}{1+z} \frac{heB_j'}{2\pi m_e c}\sigma_j^2 \\
&\sim13\,\left(\frac{1}{1+z}\right)\left(\frac{\Gamma_j}{10}\right)\left(\frac{B_j'}{1.2\times10^2\,{\rm G}}\right) 
\left(\frac{\sigma_j}{1.5\times10^4}\right)^2\,{\rm keV},
\end{aligned}
\end{equation}
where $z$ is the redshift.
This is consistent with the observed photon energies of the prompt emission from jetted TDEs. The synchrotron power radiated by one particle in the comoving frame is 
\begin{equation}
\begin{aligned}
P'_{\rm syn}&\approx \frac{4}{3}\sigma_T c \sigma_j^2 \frac{B_j'^2}{8\pi} \\
&\sim1.5\times10^{-2}\,\left(\frac{\sigma_j}{1.5\times10^4}\right)^2\left(\frac{B_j'}{1.2\times10^2\,{\rm G}}\right)^2\,{\rm erg\,s^{-1}}.
\end{aligned}
\end{equation}
The resulting prompt emission luminosity can then be estimated as
\begin{equation}
\begin{aligned}
L_{X,\rm jet}
&\approx
\Gamma_j^2 N_{\rm inj}P'_{\rm syn} \\
&\sim
2.9\times10^{46}\,\left(\frac{\Gamma_j}{10}\right)^2\left(\frac{N_{\rm inj}}{2.0\times10^{46}}\right) \\
&\:\:\:\:\:\:\:\times\left(\frac{\sigma_j}{1.5\times10^4}\right)^2\left(\frac{B_j'}{1.2\times10^2\,{\rm G}}\right)^2\,{\rm erg\,s^{-1}}. \label{Prompt luminosity}
\end{aligned}
\end{equation}
Equation~\eqref{Prompt luminosity} assumes that all particles injected at the jet base contribute to the prompt X-ray emission. In reality, only a fraction of the injected particles, plausibly $\sim1$--$10\%$, 
may contribute to the observed emission \citep[cf.][]{Sironi2015,Ball2018,Chernoglazov2023}. Even after accounting for this reduction, the predicted luminosity remains sufficient to explain the observed prompt X-ray emission. 
The magnetic-field strengths scale with the mass accretion rate as $B_H\propto B'_j\propto\dot{M}^{1/2}$, and the prompt X-ray luminosity is predicted to evolve as $L_{X,\rm jet}\propto t^{-5/3}$ because of $\dot{M}\propto t^{-5/3}$ (see Figure~\ref{fig:lightcurve}). The jet is expected to remain active until the accretion flow transitions from a geometrically thick super-Eddington disk to a geometrically thin standard disk \citep{Tchekhovskoy2014,Shen2014}. Such a transition has been associated with the abrupt X-ray shutoffs observed in several jetted TDEs at $t_{\rm off}\sim10^7$--$10^8\,{\rm s}$ \citep{Zauderer2013,Pasham2015,Eftekhari2024}.

The CDI can generate large-scale current sheets within the jet. Assuming the characteristic size of the reconnection layer is comparable to the jet cylinder radius, the duration of a reconnection event in the jet comoving frame can be estimated as \citep{Sironi2025}
\begin{equation}
t'_{\rm rec}
\sim
\frac{R_j}{\beta_{\rm rec}c}.
\end{equation}
At the characteristic dissipation radius, the jet cylinder radius is $R_j\sim30\,r_g$ for a parabolic jet ($\alpha=1/2$). The number density of the jet plasma at the dissipation radius is estimated from the continuity equation as
$n_{\pm,\rm dis}\approx (r_g/R_j)^2 n_{\pm}\sim10^{-3}n_{\pm}$,
suggesting that the plasma is collisionless at the dissipation radius (see Eq.~\eqref{collision length scale}). 
The observed duration of the flare is
\begin{equation}
\begin{aligned}
\delta t_{\rm obs}
&\sim \frac{(1+z)t'_{\rm rec}}{\Gamma_j} \\
&\sim 150\,(1+z)
\left(\frac{R_j/r_g}{30}\right)
\left(\frac{M_{\rm BH}}{10^6\,M_\odot}\right)\\
&\:\:\:\:\:\:\:\times\left(\frac{\beta_{\rm rec}}{0.1}\right)^{-1}\left(\frac{\Gamma_j}{10}\right)^{-1}\,{\rm s}.
\end{aligned}
\end{equation}
This is consistent with the observed variability.

Finally, we examine whether the prompt X-ray photons can escape the emission region without significant electron scattering. Using the plasma density at the emission region, the Thomson scattering mean free path is $l_T = 1/(n_{\rm \pm,\rm dis} \sigma_T) \approx 10^{15}\,{\rm cm}$, which exceeds the dissipation radius $r_{\rm dis}\approx 10^{14}\,{\rm cm}$. A conservative estimate of the local scattering optical depth is $\tau_T \lesssim  r_{\rm dis}/l_{T} \approx 0.07$, indicating that the prompt emission region is optically thin to electron scattering.

\section{Detectability of MeV/GeV gamma-ray emission}
\label{sec:discussion}

\begin{figure}
\centering
\includegraphics[width=0.45\textwidth]{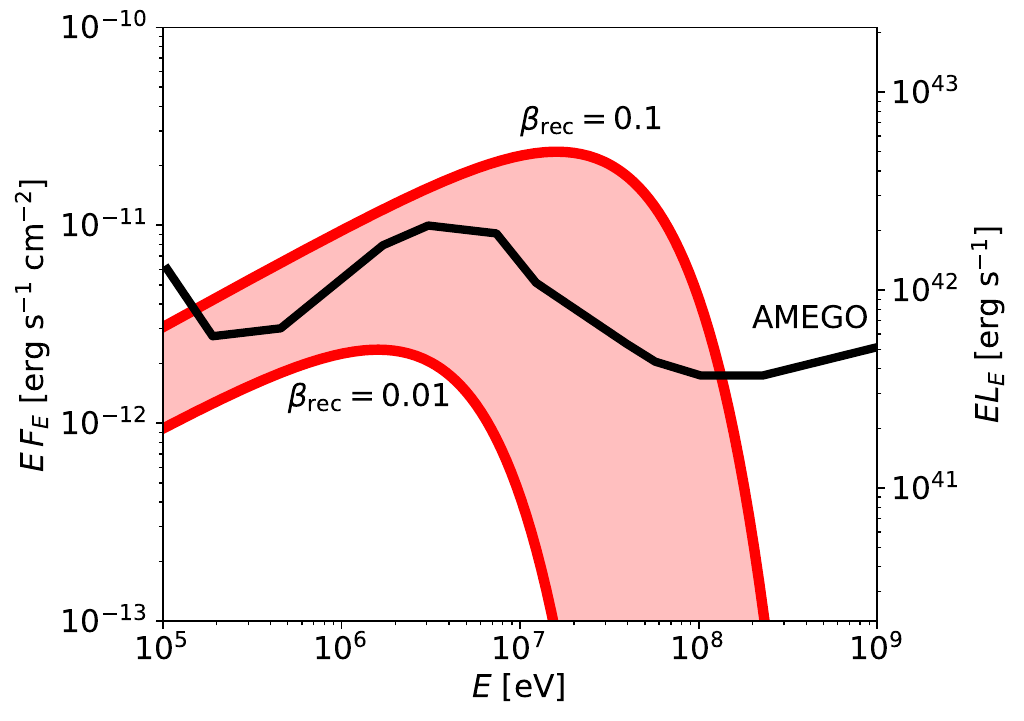}
\vspace{-0.05cm}
\caption{
Predicted flux from magnetic reconnection in the vicinity of the BH for a TDE at $z \sim 0.01$. 
The red shaded regions correspond to reconnection rates of $\beta_{\rm rec}=0.1$ (collisionless regime; upper) and $\beta_{\rm rec}=0.01$ (collisional regime; lower), respectively. 
The black curve shows the sensitivity of AMEGO for an exposure time of $10^6$~s \citep{AMEGO2019}. 
The predicted MeV gamma-ray emission is expected to precede the prompt emission (see Figure~\ref{fig:lightcurve}).
}
\label{fig:SED}
\end{figure}
\begin{figure}
\centering
\includegraphics[width=0.45\textwidth]{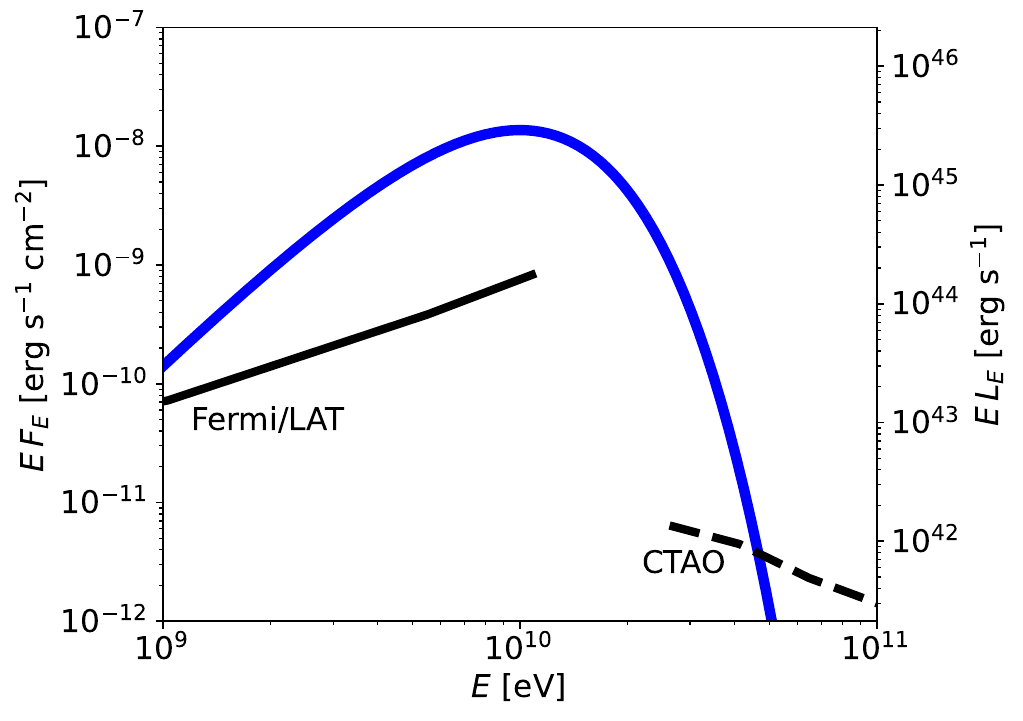}
\vspace{-0.05cm}
\caption{
Predicted EIC flux of GeV gamma-rays for a jetted TDE at $z \sim 0.01$. 
The blue line shows the EIC emission produced by inverse-Compton scattering of thermal photons from a super-Eddington accretion disk by the electrons responsible for the prompt X-ray emission. 
The black solid and dashed lines represent the sensitivities of {\it Fermi}/LAT and CTAO, respectively, for an exposure time of $10^5$~s. 
Since the EIC emission is produced by the same electrons that power the prompt emission, it is expected to be approximately contemporaneous with the prompt emission.
}
\label{fig:SED_GeV}
\end{figure}

We now discuss the detectability of the high-energy emission predicted by the reconnection-driven plasma loading scenario. Our scenario predicts two high-energy components with distinct physical origins and temporal signatures: a MeV gamma-ray synchrotron component produced by magnetic reconnection near the BH before the onset of the prompt emission, and a contemporaneous GeV gamma-ray component generated by inverse-Compton scattering of thermal disk photons by prompt emission electrons. 
We compare the predicted fluxes with the sensitivities of the All-sky Medium Energy Gamma-ray Observatory (AMEGO) \citep{AMEGO2019}, the {\it Fermi} Large Area Telescope ({\it Fermi}/LAT) \citep{LAT2009}, and the Cherenkov Telescope Array Observatory (CTAO) \citep{CTA2021}.

One prediction of our scenario is MeV gamma-ray emission produced by magnetic reconnection near the BH.
Magnetic reconnection accelerates electrons to relativistic energies, which produce synchrotron emission extending into the MeV gamma-ray band (see Section~\ref{sec:radiation_reconnection}).
In our scenario, a fraction of these photons undergo pair production at the jet base. However, Eq~\eqref{optical depth} indicates that most of the photons escape from the system without interacting, suggesting that the MeV gamma rays responsible for plasma loading may be detectable.
We consider reconnection rates in the range $0.01\lesssim\beta_{\rm rec}\lesssim0.1$, encompassing the values expected in both collisional and collisionless reconnection regimes. We then calculate the resulting emission spectrum using the characteristic synchrotron photon energy and reconnection luminosity given by Eqs~(\ref{eq:E_syn}) and (\ref{eq:L_rec}), respectively.
As shown in Figure~\ref{fig:SED}, the reconnection component peaks in the MeV band and can approach the sensitivity of AMEGO for a nearby jetted TDE at $z\sim0.01$. This emission is produced before the onset of the prompt emission. Its detection would provide a direct test of the reconnection-driven plasma loading scenario.

A second prediction is GeV gamma-ray emission from the prompt emission region, produced when the electrons responsible for our prompt emission upscatter thermal disk photons through the external inverse Compton (EIC) process \citep{Lu2016}.
Because the EIC emission is produced by the same prompt emission electrons, it is expected to be approximately contemporaneous with the prompt emission.
In a TDE involving a main-sequence star, optically thick super-Eddington accretion flow is expected to form, with its emission peaking at
$E_{\rm seed}\sim 100~{\rm eV}\left({T}/{10^6~{\rm K}}\right).$
In the Thomson regime, EIC scattering of these photons by the prompt emission electrons produces photons with a characteristic energy given by
\begin{equation}
E_{\rm EIC}\sim \sigma_j^2 E_{\rm seed}
\sim 10~{\rm GeV}
\left(\frac{\sigma_j}{10^4}\right)^2
\left(\frac{E_{\rm seed}}{100~{\rm eV}}\right).
\end{equation}
At the prompt emission region, the photon and magnetic energy densities are estimated as
\begin{equation}
\begin{aligned}
U_{\rm ph}&=\frac{L_{\rm{d}}}{4\pi r_{\rm em}^2c}\\
&\sim 3\times10^3
\left(\frac{L_{\rm d}}{10^{43}~{\rm erg\,s^{-1}}}\right)
\left(\frac{r_{\rm em}}{10^{14}~{\rm cm}}\right)^{-2}
~{\rm erg\,cm^{-3}},
\end{aligned}
\end{equation}
and
\begin{equation}
U_B=\frac{B_j^2}{8\pi}\sim4\times10^4
\left(\frac{B_j}{10^3~{\rm G}}\right)^2
~{\rm erg\,cm^{-3}},
\end{equation}
respectively, where $L_{\rm d}$ is the disk luminosity. For the fiducial parameters adopted above, the peak EIC luminosity is therefore estimated to be
\begin{equation}
\begin{aligned}
L_{\rm EIC}&\sim\frac{U_{\rm ph}}{U_B}L_{\rm X,jet}\\
&\sim 10^{44}\,\left(\frac{U_{\rm ph}/U_B}{0.1}\right)\left(\frac{L_{\rm X,jet}}{10^{45}\,\rm erg\, s^{-1}}\right)\,{\rm erg\, s^{-1}}. 
\end{aligned}
\end{equation}
Figure~\ref{fig:SED_GeV} shows the predicted EIC spectrum for a nearby jetted TDE at $z\sim0.01$, together with the sensitivities of {\it Fermi}/LAT and CTAO.
The predicted GeV emission may be detectable by {\it Fermi}/LAT and CTAO for a nearby event at $z\sim0.01$, providing an additional means of testing our scenario.
Future observations in the MeV and GeV gamma-ray bands are expected to provide a test of the reconnection-driven plasma loading scenario for BH jets.

\section{Summary \& Discussion}
\label{Sec:summary}

We have investigated magnetic reconnection in the vicinity of the BH and the subsequent production of electron--positron pairs as a plasma loading mechanism for TDE jets. We have shown that magnetic reconnection can be triggered near the BH within a super-Eddington accretion flow and that the resulting gamma rays can generate copious electron--positron pairs. The injected pair density at the jet base reaches $\sim10^{10}$ times the Goldreich--Julian density, and the particle supply is sufficient to account for the observed prompt emission.
In our scenario, the jet remains electromagnetically dominated out to the prompt emission region. The CDI is expected to generate large-scale current sheets within the jet, where magnetic reconnection dissipates magnetic energy and produces the prompt emission. We have shown that the characteristic emission radius, photon energy, luminosity, variability timescale predicted by CDI triggered magnetic reconnection are consistent with the observed properties of jetted TDEs. In our scenario, an MeV gamma-ray signals from magnetic reconnection near the BH and a GeV gamma-ray signals from the prompt emission region are predicted. Current and future observations with AMEGO, {\it Fermi}/LAT, and CTAO will provide direct tests of our scenario.

Regarding the radiation mechanism, our plasma loading scenario suggests that the prompt emission is produced by synchrotron radiation. One of the alternative mechanisms is photospheric emission, in which thermal photons escape from the photosphere \citep{Goodman1986,peter2006,Crumley2016}. However, if the prompt emission is photospheric, Eq.~\eqref{Required number particles} cannot be directly applied to estimate the required number of particles in the jet. To explain the prompt emission through photospheric emission at the same emission radius $R_{\rm dis}$, the emission region must be located near or below the photosphere. The optical depth to electron scattering must satisfy
\begin{equation}
\tau_{\rm ph}
\approx n_{\pm,\rm dis}\sigma_T\frac{R_{\rm dis}}{\Gamma_j}
\sim{1}.
\end{equation}
The required number density is therefore at least
\begin{equation}
\begin{aligned}
n_{\pm,\rm dis}
&\approx \frac{\Gamma_j}{\sigma_T R_{\rm dis}} \\
&\sim1.5\times10^{11}
\left(\frac{\Gamma_j}{10}\right)
\left(\frac{R_{\rm dis}}{10^{14}\,{\rm cm}}\right)^{-1}
\,{\rm cm}^{-3}.
\end{aligned}
\end{equation}
This density is more than two orders of magnitude higher than that predicted by our plasma injection scenario. Supplying such a large amount of plasma is not addressed in our current model and would require a different plasma-loading mechanism. How the conditions required to trap photons within TDE jets are established is therefore an open question for future investigation.
The detection of GeV gamma-ray emission from the prompt emission region, as discussed in Section~\ref{sec:discussion}, may help distinguish between photospheric and synchrotron interpretations of the prompt emission. In the photospheric model, Comptonization of the thermal spectrum {can} produce a high-energy tail in the X-ray band \citep{peter2006}. The Lorentz factors of the electrons involved in this process are $\sim10^2$ \citep{Crumley2016}, much smaller than the value of $\sim10^4$ inferred in our scenario. Consequently, these electrons are unlikely to produce GeV gamma-ray emission through EIC scattering. Thus, a future detection of GeV gamma-ray emission from the prompt emission region could help distinguish the specific photospheric scenario involving lower-energy electrons from our synchrotron scenario. Conversely, a sufficiently sensitive nondetection of the GeV gamma-ray emission predicted by our model could provide relative support for a photospheric interpretation. In that case, the origin of the plasma required to trap photons within the TDE jet would need to be investigated further.

\begin{acknowledgements}

We thank 
Itai~Linial,
Hirofumi~Noda,
Kenji Toma, and
Nimrod~Tripto
for valuable comments.
We also thank the anonymous referee for their helpful comments that improved the paper.
R.O. was supported by the Graduate Program on Physics for the Universe (GP-PU).
This research was partially supported by JSPS KAKENHI Grant No. 25KJ0010 (Y.S.).

\end{acknowledgements}

\begin{contribution}
The authors contributed equally to this work.
\end{contribution}

\bibliography{sample701}{}
\bibliographystyle{aasjournalv7}

\end{document}